\documentclass[english]{article}
\usepackage[T1]{fontenc}
\usepackage[latin1]{inputenc}
\usepackage{geometry}
\usepackage{graphicx}
\usepackage{setspace}
\makeatletter

\usepackage{amsmath,amssymb}
\date{}

\usepackage{babel}
\makeatother
\begin{document}
\begin{onehalfspace}
\begin{flushright}
{\tt hep-ph/yymmdd}\\
ITP Budapest 608\\
December 2003\\
\end{flushright}
\smallskip{}
\end{onehalfspace}

\begin{center}{\LARGE Vector Condensate Model of Electroweak Interactions}\end{center}{\LARGE \par}
\smallskip{}

\begin{onehalfspace}
\begin{center}{\large G. Cynolter$^{*}$, E. Lendvai$^{*}$ and G.
Pócsik$^{\dagger}$ }\end{center}{\large \par}
\smallskip{}

\begin{center}\textit{$^{*}$Theoretical Physics Research Group of
Hungarian Academy of Sciences, Eötvös University, Budapest, Hungary}\\
\textit{$^{\dagger}$Institute for Theoretical Physics, Eötvös Lorand
University, Budapest, Hungary }\end{center}
\smallskip{}
\end{onehalfspace}

\begin{abstract}
In the standard model of electroweak interactions the Higgs doublet
is replaced by a complex vector doublet and a real vector singlet.
The gauge symmetry is broken dynamically by a mixed condensate of
the doublet and singlet vector fields. Gauge fields get their usual
standard model masses by condensation. The new vector matter fields
become massive by their gauge invariant selfcouplings. Fermions are
assigned to the gauge group in the usual manner. Fermion masses are
coming from a gauge invariant fermion-vector field interaction by
a mixed condensate. The Kobayashi-Maskawa description is unchanged.
It is shown that from the new matter fields a large number of spin-one
particle pairs is expected at future high energy $e^{+}e^{-}$ linear
colliders of 500-1500 GeV.
\end{abstract}
A current description of electroweak symmetry breaking is through
a weakly interacting scalar doublet. Another possibility is a symmetry
breaking system interacting strongly with the longitudinal weak vector
bosons which has been realised in the DHT model$^{1}$ based on a
chiral Lagrangian approach. An alternative description of the strongly
interacting symmetry breaking system has been proposed in the BESS
model$^{2}$ through nonlinear realisations. Top quark condensation
has also been suggested for describing the electroweak symmetry breaking$^{3}$
leading to several interesting studies (e.g.Ref.4). Electroweak symmetry
breaking caused by the condensation of a vector field was studied,
too$^{5}$. Condensation of vector bosons in different scenarios was
considered in the literature$^{6}$. Recently, little Higgs models$^{7}$
attracted attention.

In the present note we start with the usual Lagrangian of the standard
model of electroweak interactions but instead of the scalar doublet
two new matter fields are introduced. One of them is a Y = 1, T =
1/2 doublet of complex vector fields \begin{equation}
B_{\mu}=\left(\begin{array}{c}
B_{\mu}^{(+)}\\
B_{\mu}^{(0)}\end{array}\right),\label{eq:bdub}\end{equation}
the other is a real Y = 0, T = 0 vector field $C_{\mu}$. This extends
our recent model$^{5}$ where only $B_{\mu}$ was present with the
condensation of $B_{\mu}^{(0)}$. Consequently, we are able to describe
a more complete symmetry breaking and to generate fermion masses from
a gauge invariant interaction Lagrangian while the mass ratio of $B^{(+)}$
and $B^{(0)}$ does not become fixed. The key point is the introduction
of a mixed $B_{\mu}-C_{\mu}$ condensate together with suitable gauge
invariant interactions of the new matter fields. This leads to nonvanishing
standard model particle masses, as well as B, C particle masses. It
turns out that altogether three condensate emerge but only one combination
of theirs is fixed by the Fermi coupling constant. The model should
be considered as a low energy effective one. Based on recent experience$^{5}$,
probably it has a few TeV cutoff scale. Its' new particle content
is a charged vector boson pair and three neutral vector bosons. As
is shown, these can be pair produced in $e^{+}e^{-}$ annihilation,
and at future linear colliders of 500-1500 GeV they can provide a
large number of events.

To build the model, in the Lagrangian of the standard model the interactions
of the scalar doublet are replaced by the gauge invariant Lagrangian 

\begin{eqnarray}
L_{BC} & = & -\frac{1}{2}\overline{\left(D_{\mu}B_{\nu}-D_{\nu}B_{\mu}\right)}\left(D^{\mu}B^{\nu}-D^{\nu}B^{\mu}\right)-\nonumber \\
 &  & -\frac{1}{2}\left(\partial_{\mu}C_{\nu}-\partial_{\nu}C_{\mu}\right)\left(\partial^{\mu}C^{\nu}-\partial^{\nu}C^{\mu}\right)-V(B,C),\label{eq:Lagrangian}\end{eqnarray}
 where $D_{\mu}$ is the covariant derivative, $g_{\mu\nu}=+---$,
and for the potential $V(B,C)$ we assume \begin{equation}
V(B,C)=\lambda_{1}\left(\overline{B}_{\nu}B^{\nu}\right)^{2}+\lambda_{2}\left(C_{\nu}C^{\nu}\right)^{2}+\lambda_{3}\overline{B}_{\nu}B^{\nu}C_{\mu}C^{\mu},\label{eq:vbc}\end{equation}
 depending only on B-, C- lengths. Other quartic terms would not change
the argument. $\lambda_{1,2,3}$ are real and from positivity \begin{equation}
\lambda_{1}>0,\quad\lambda_{2}>0,\;\quad4\lambda_{1}\lambda_{2}>\lambda_{3}^{2}.\label{eq:lambdarel}\end{equation}
 Mass terms to be generated are not introduced explicitely in (\ref{eq:vbc}).
Fermion-BC interactions are introduced later on. 

To break the gauge symmetry, we assume a nonvanishing mixed condensate
in the vacuum ,\begin{equation}
\left\langle \overline{B}_{\mu}C_{\nu}\right\rangle =g_{\mu\nu}\:\:(0,d),\label{eq:5}\end{equation}
 where the left-hand side could be rotated into (0,d), $d\neq0$,
respecting also electric charge conservation and defining the neutral
and charged components in (\ref{eq:bdub}). By $U_{Y}(1)$ $d$ can
be chosen real. A real $d$ respects also combined TCP and C symmetries.
By TCP-invariance (5) equals $\left\langle C_{\nu}\overline{B}_{\mu}\right\rangle $.
It follows from (5) that the only nonvanishing mixed condensate is
\begin{equation}
\left\langle B_{1\mu}C_{\nu}\right\rangle =\sqrt{2}g_{\mu\nu}d\label{eq:6}\end{equation}
with\begin{equation}
B_{\mu}^{(0)}=\frac{{1}}{\sqrt{{2}}}\left(B_{1\mu}+iB_{2\mu}\right),\label{eq:7}\end{equation}
where $B_{j\mu}$ is real. Once there exists the mixed condensate,
most B and C condense separately, too, therefore, we assume \begin{eqnarray}
\left\langle \overline{B}_{\mu}B_{\nu}\right\rangle  & = & g_{\mu\nu}k_{1},\quad k_{1}\neq0,\label{eq:8}\\
\left\langle C_{\mu}C_{\nu}\right\rangle  & = & g_{\mu\nu}k_{3},\quad k_{3}\neq0.\nonumber \end{eqnarray}
In more detail, assume that $k_{1}$ originates from $B_{\mu}^{(0)}$
condensation, \begin{eqnarray}
\left\langle B_{\mu}^{(+)\dagger}B_{\nu}^{(+)}\right\rangle  & = & 0,\nonumber \\
\left\langle B_{\mu}^{(0)\dagger}B_{\nu}^{(0)}\right\rangle  & = & g_{\mu\nu}k_{1}.\label{eq:9}\end{eqnarray}
(9) reproduces the pattern of gauge particle masses$^{5}$. All the
condensates linear in $B_{\mu}^{(+)}$ vanish by charge conservation.
Finally, we assume in advance, that \begin{equation}
\left\langle B_{\mu}^{(0)}B_{\nu}^{(0)}\right\rangle =\left\langle B_{\mu}^{(0)\dagger}B_{\nu}^{(0)\dagger}\right\rangle =g_{\mu\nu}\:\: k_{2}.\label{eq:10}\end{equation}
The point is that in general $B_{1\mu}$ and $B_{2\mu}$ belong to
different masses, so that $k_{2}\neq0$. $k_{1,2,3}$ are real and
$k_{1}<0$, $k_{3}<0$, as shown by particle masses and simple models.
The condensates are of nonperturbative origin caused by the strong
interaction V(B,C). Among them only $k_{1}$ is fixed by contemporary
phenomenology.

Mass terms are obtained in the linearized form of $L_{BC}$ via condensates.
The $W^{\pm}$ mass is determined by the total B-condensate, while
the two neutral gauge field combinations are proportional to $B_{\mu}^{(+)\dagger}B_{\nu}^{(+)}$
and $B_{\mu}^{(0)\dagger}B_{\nu}^{(0)}$, respectively. Therefore,
the assumption (9) yields\begin{equation}
m_{\textrm{photon }}=0,\; m_{W}=\frac{g}{2}\sqrt{-6k_{1}},\; m_{Z}=\frac{g}{2\cos\theta_{W}}\sqrt{-6k_{1}}.\label{eq:mwz}\end{equation}
 Low energy phenomenology gives\begin{equation}
k_{1}=-\left(6\sqrt{2}G_{F}\right)^{-1},\;\left(-6k_{1}\right)^{1/2}=246\textrm{ GeV}.\label{eq:k1}\end{equation}
$B^{\pm}$ and $B_{2}$ get the following masses\begin{eqnarray}
m_{\pm}^{2} & = & -8\lambda_{1}k_{1}-4\lambda_{3}k_{3},\label{eq:mpm}\\
m_{B_{2}}^{2} & = & -10\lambda_{1}k_{1}+2\lambda_{1}k_{2}-4\lambda_{3}k_{3}=m_{\pm}^{2}+2\lambda_{1}(k_{2}-k_{1)}.\nonumber \end{eqnarray}
 For $\lambda_{3},-k_{1},-k_{3}>0$, $\; m_{B_{2}}^{2}>m_{\pm}^{2}>0$
since $k_{2}>k_{1}$. The $B_{1}-C$ sector is slightly more complicated,
here one arrives at the following bilinear combinations in the potential
for $B_{1\mu},C_{\mu}$\begin{equation}
V(B,C)\rightarrow-\frac{m_{1}^{2}}{2}B_{1\mu}B^{1\mu}-\frac{m_{2}^{2}}{2}C_{\nu}C^{\nu}-m_{3}^{2}B_{1\mu}C^{\mu},\label{eq:14}\end{equation}
 with \begin{eqnarray}
-m_{1}^{2} & = & 10\lambda_{1}k_{1}+2\lambda_{1}k_{2}+4\lambda_{3}k_{3}=-m_{B_{2}}^{2}+4\lambda_{1}k_{2},\nonumber \\
-m_{2}^{2} & = & 24\lambda_{2}k_{3}+8\lambda_{3}k_{1},\label{eq:m123}\\
-m_{3}^{2} & = & 4\sqrt{2}\lambda_{3}d.\nonumber \end{eqnarray}
 Here $m_{1}^{2}>0$ being $k_{1}+k_{2}<0$; $m_{2}^{2}>0$$m_{3}^{2}\lessgtr0$.
A positive potential in (14) requires\begin{equation}
m_{1}^{2},m_{2}^{2}>0,\;\quad m_{1}^{2}m_{2}^{2}>m_{3}^{4}.\label{eq:16}\end{equation}

(\ref{eq:14}) shows that $B_{1\mu}$ and $C_{\mu}$are nonphysical
fields, the mass eigenstates are defined by\begin{eqnarray}
B_{f\mu} & = & cB_{1\mu}+sC_{\mu},\nonumber \\
C_{f\mu} & = & -sB_{1\mu}+cC_{\mu},\label{eq:mphys}\end{eqnarray}
 where $c=\cos\phi,\; s=\sin\phi,$ $\phi$ denotes the mixing angle
defined by \begin{equation}
\frac{1}{2}\sin2\phi(m_{1}^{2}-m_{2}^{2})=\cos2\phi m_{3}^{2}.\label{eq:18}\end{equation}
 The physical masses are\begin{eqnarray}
m_{B_{f}}^{2} & = & c^{2}m_{1}^{2}+s^{2}m_{2}^{2}+2csm_{3}^{2},\nonumber \\
m_{C_{f}}^{2} & = & s^{2}m_{1}^{1}+c^{2}m_{2}^{2}-2csm_{3}^{2},\label{eq:mbc}\end{eqnarray}
whence\begin{eqnarray}
2m_{B_{f},C_{f}}^{2} & = & m_{1}^{2}+m_{2}^{2}\pm\frac{m_{1}^{2}-m_{2}^{2}}{\cos2\phi}.\label{eq:m12}\end{eqnarray}
 For $(m_{1}^{2}-m_{2}^{2})/\cos2\phi>0$ ($<0$) $m_{B_{f}}^{2}>m_{C_{f}}^{2}>0$
($m_{C_{f}}^{2}>m_{B_{f}}^{2}>0$). At vanishing mixing, $m_{3}^{2}=0$,
$B_{1\mu}$ and $C_{\mu}$ become independent having the masses $m_{1}$
and $m_{2}$; taking $k_{2}=0$ and omitting $C_{\mu}$ we recover
the model of Ref.5. $k_{2}$ shifts the real component field masses
from the mass of the imaginary part $B_{2\mu}$.

The particle spectrum of the B-C sector consists of the spin-one $B^{\pm}$
and the three neutral spin-one particles $B_{2},B_{f},C_{f}$. Their
masses are rather weakly restricted. Beside the gauge coupling constants
and $\lambda_{1},\lambda_{2},\lambda_{3}$, the model has three basic
condensates $\left\langle V_{i\mu}V_{i\nu}\right\rangle $, $V_{i\mu}=B_{2\mu},B_{f\mu},C_{f\mu}$.
$k_{1},k_{2},k_{3},d$ condensates are built up from these as follows\begin{eqnarray}
g_{\mu\nu}d & = & \frac{1}{\sqrt{2}}cs\left(\left\langle B_{f\mu}B_{f\nu}\right\rangle -\left\langle C_{f\mu}C_{f\nu}\right\rangle \right),\nonumber \\
g_{\mu\nu}k_{1} & = & \frac{1}{2}\left\{ c^{2}\left\langle B_{f\mu}B_{f\nu}\right\rangle +s^{2}\left\langle C_{f\mu}C_{f\nu}\right\rangle +\left\langle B_{2\mu}B_{2\nu}\right\rangle \right\} ,\nonumber \\
g_{\mu\nu}k_{2} & = & \frac{1}{2}\left\{ c^{2}\left\langle B_{f\mu}B_{f\nu}\right\rangle +s^{2}\left\langle C_{f\mu}C_{f\nu}\right\rangle -\left\langle B_{2\mu}B_{2\nu}\right\rangle \right\} \label{eq:k123d}\\
g_{\mu\nu}k_{3} & = & s^{2}\left\langle B_{f\mu}B_{f\nu}\right\rangle +c^{2}\left\langle C_{f\mu}C_{f\nu}\right\rangle .\nonumber \end{eqnarray}
 From (\ref{eq:k123d}) $d$ can be written as \begin{equation}
2\sqrt{2}\cot2\phi\: d=k_{1}+k_{2}-k_{3}.\label{eq:d-k}\end{equation}

Turning to the dynamical fermion mass generation, we add to the gauge
vector and matter vector field Lagrangians, the usual fermion-gauge
vector Lagrangian, as well as a new gauge invariant piece responsible
for the fermion-matter vector field interactions and in usual notation
this is (for quarks)\begin{eqnarray}
g_{ij}^{u}\overline{\psi}_{iL}u_{jR}B_{\nu}^{C}C^{\nu}+g_{ij}^{d}\overline{\psi}_{iL}d_{jR}B_{\nu}c^{\nu}+h.c.,\label{eq:mfermion}\\
\psi_{iL}=\left(\begin{array}{c}
u_{i}\\
d_{i}\end{array}\right)_{L},\quad B_{\nu}^{C}=\left(\begin{array}{c}
B_{\nu}^{(0)\dagger}\\
-B_{\nu}^{(+)\dagger}\end{array}\right).\nonumber \end{eqnarray}

Clearly the mixed condensate provides fermion masses and also the
Kobayashi-Maskawa description is unchanged. A typical fermion mass
is\begin{equation}
m_{f}=-4g_{f}d\label{eq:mfermd}\end{equation}
and only $g_{f}d$ becomes fixed but $m_{f1}/m_{f2}=g_{f1}/g_{f2}$
as usual. If $d$ is about $k_{1}\simeq G_{F}^{-1}$, then $g_{f}$
is a factor of $G_{F}^{1/2}$ weaker than the approximate standard
model value $G_{F}^{1/2}$.

As for the interactions of the new vector particles, at present we
confine ourselves only to a few remarks. There exist $V_{i}V_{j}V-$,$V_{i}V_{j}VV-$
type couplings with gauge bosons $V=\gamma,W^{\pm},Z$, $\overline{f}_{1}f_{2}V_{i}V_{j}$
-type interactions and $V_{i}V_{j}V_{k}V_{l}$-type matter vector
couplings always with an even number of $V_{i}$. Depending on the
mass hierarchy, one or more $V_{i}$ may be stable. Relatively large
are the $V_{i}V_{j}V$ couplings.

As an example we consider the $Z-B_{f}- B_{2}$ coupling, \begin{eqnarray}
L_{I} & = & \frac{g}{2\cos\theta_{W}}\cos\phi\cdot Z_{\mu}\left[B_{f\nu}(\partial^{\mu}B_{2}^{\nu}-\partial^{\nu}B_{2}^{\mu})-B_{2\nu}(\partial^{\mu}B_{f}^{\nu}-\partial^{\nu}B_{f}^{\mu})\right].\label{eq:Lint}\end{eqnarray}
Direct production of $B_{f}B_{2}$ pairs can be studied in high energy
$e^{+}e^{-}$ colliders, $e^{+}e^{-}\rightarrow Z^{*}\rightarrow B_{f}B_{2}$.
Assume in (\ref{eq:mfermd}) $g_{e^{-}}$ is very small, then the
direct $e^{+}e^{-}\rightarrow B_{f}B_{2}$ can be neglected. From
(\ref{eq:Lint}) we have for the total cross section\[
\sigma(e^{+}e^{-}\rightarrow Z^{*}\rightarrow B_{f}B_{2})=\frac{g^{4}\cos^{2}\phi}{3\cdot4096\cos^{4}\theta_{W}}\frac{1+(4\sin^{2}\theta_{W}-1)^{2}}{m_{B_{2}}^{2}m_{B_{f}}^{2}s^{2}(s-M_{Z}^{2})^{2}}\left(s-(m_{B_{2}}+m_{B_{f}})^{2}\right)^{3/2}\cdot\]
\begin{equation}
\cdot\left(s-(m_{B_{f}}-m_{B_{2}})^{2}\right)^{3/2}\left(2s(m_{B_{2}}^{2}+m_{B_{f}}^{2})+m_{B_{f}}^{4}+m_{B_{2}}^{4}+10m_{B_{2}}^{2}m_{B_{f}}^{2}\right).\label{eq:sigma}\end{equation}

At asymptotic energies $\sigma$ is proportional to $1/m_{B_{2}}^{2}+1/m_{B_{f}}^{2}$.
The mass and energy dependences of $\sigma$ are shown in Figs. 1,2.
For example at $\sqrt{s}$= 500 GeV and with an integrated luminosity
of 10 fb$^{-1}$ 5700, 1900, 530 $B_{f}B_{2}$ pairs are expected
for $m_{B_{f}}=m_{B_{2}}$= 100, 150, 200 GeV and $\cos^{2}\phi=1/2$.
At $\sqrt{s}$ = 1.5 TeV a higher mass range can be tested, for $\cos^{2}\phi=1/2$,
a luminosity of 100 fb$^{-1}$ we get the large event numbers 62200,
14500, 5900, 1900, 530 for $m_{B_{f}}=m_{B_{2}}$= 100, 200, 300,
400, 500 GeV. One can show that the $B^{+}B^{-}$ production is a
factor of $\cos^{2}2\theta_{W}$ smaller than (\ref{eq:sigma}) at
equal masses and $\cos^{2}\phi=1$.

In conclusion, a low energy dynamical symmetry breaking model of electroweak
interactions based on matter vector field condensation is introduced.
Mass generation is arranged starting from gauge invariant Lagrangians.
New particles are all spin-one states, one charged pair and three
neutral particles having many interactions. The parameter space of
the model is larger than that of the one in Ref.5, therefore, we expect
that the positive result of the S,T parameter analysis can be maintained.
We hope to investigate the model further in a future work.

This work is supported in part by OTKA, No. T029803.

\newpage
\section*{Figures}

\includegraphics[%
  scale=0.5,
  angle=270]{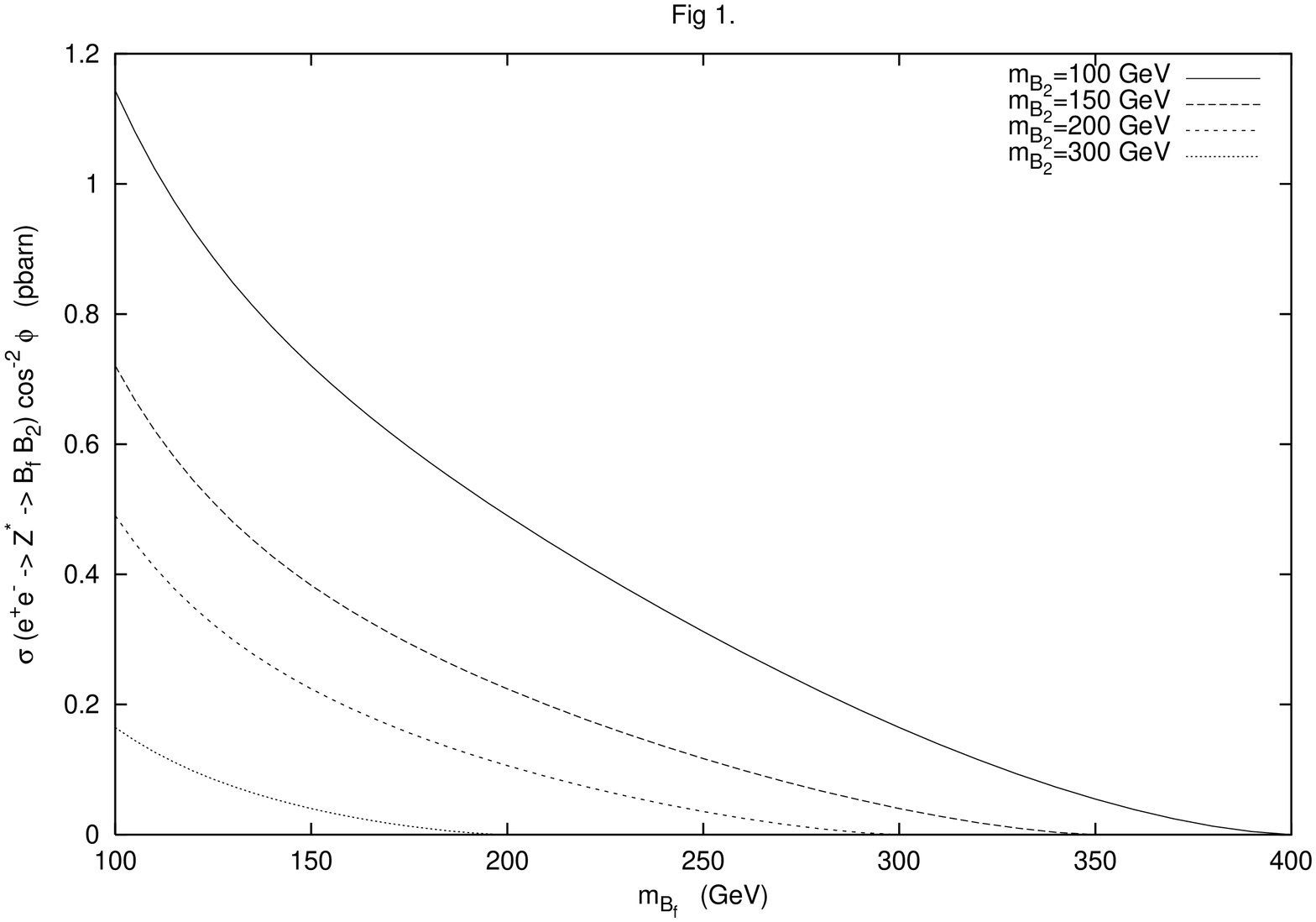}

Fig1. $\cos^{-2}\phi\;\sigma\left(e^{+}e^{-}\rightarrow B_{f}B_{2}\right)$
vs. $m_{B_{f}}$at $\sqrt{s}=500$ GeV and various $m_{B_{2}}.$

\includegraphics[%
  scale=0.5,
  angle=270]{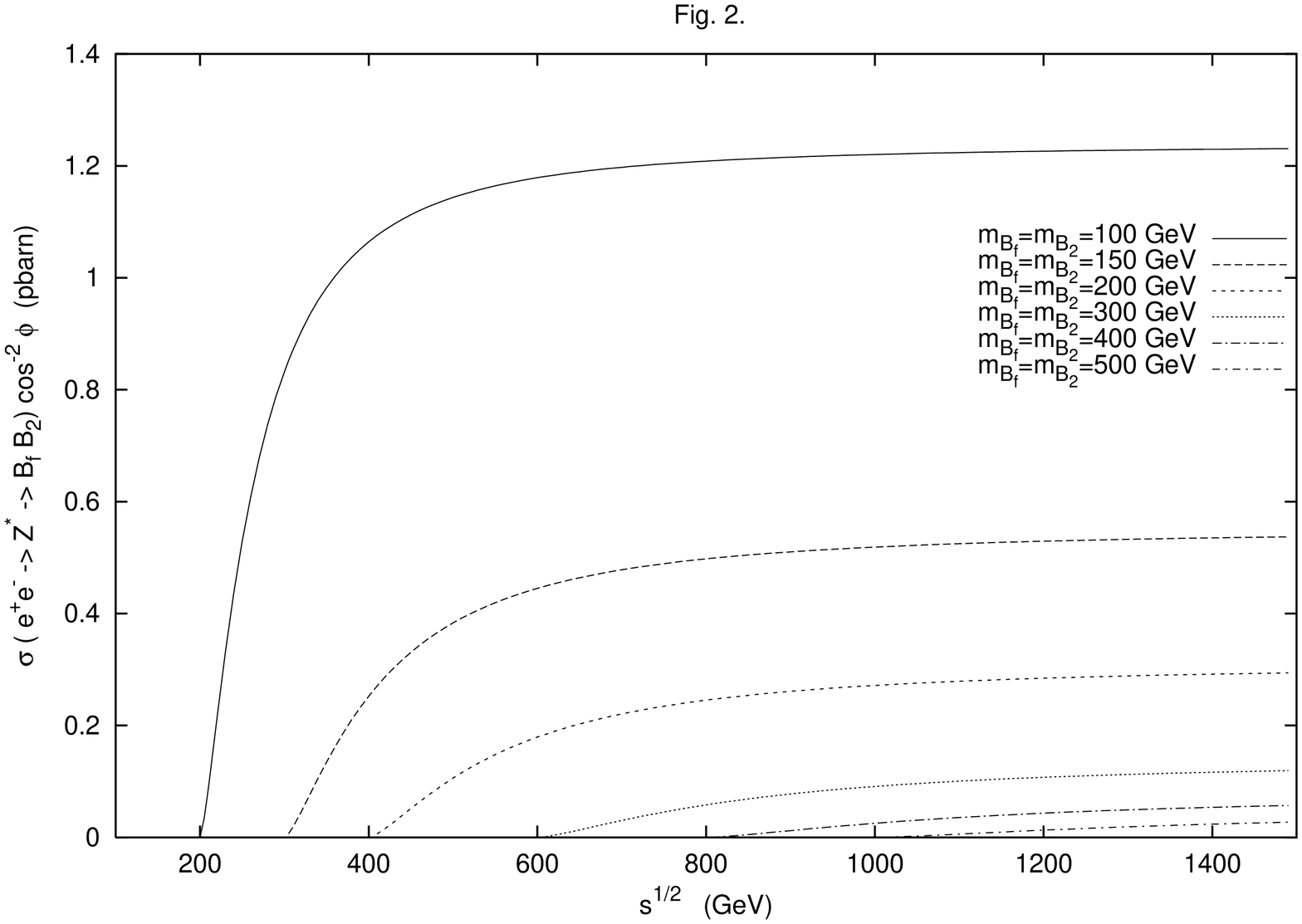}

Fig2. $\cos^{-2}\phi\;\sigma\left(e^{+}e^{-}\rightarrow B_{f}B_{2}\right)$
vs. $\sqrt{s}$ at various $m_{B_{f}}=m_{B_{2}}$.
\end{document}